\documentclass[grl]{agu2001}

\usepackage{graphicx}
\usepackage{dcolumn}
\usepackage{bm}
\usepackage{amsmath,amssymb,float,alltt}
\usepackage{color}

\authorrunninghead{LAPENTA ET AL.}
\titlerunninghead{Signatures of Magnetic Reconnection }

\journalid{????} \articleid{???}{??} \paperid{????}
\cpright{AGU}{2007}

\received{????} \revised{????} \accepted{????} \published{}

\authoraddr{G. Lapenta, S. Markidis, A. Divin, Centrum voor Plasma-Astrofysica, Departement Wiskunde, Katholieke Universiteit Leuven, Celestijnenlaan 200B, 3001 Leuven, Belgium (giovanni.lapenta@wis.kuleuven.be)}
\authoraddr{M. Goldman, D. Newman, Department of Physics, University of Colorado}

\newcommand{\bfE}{\mathbf{E}}
\newcommand{\bfB}{\mathbf{B}}

\begin{document}

\title{Bipolar Electric Field Signatures of Reconnection Separatrices for a Hydrogen Plasma at Realistic Guide Fields}

\author{G. Lapenta, S. Markidis, A. Divin}
\affil{Centrum voor Plasma-Astrofysica,
Departement Wiskunde, Katholieke Universiteit Leuven,
Belgium (EU).}
\author{M. Goldman, D. Newman}
 \affil{Department of Physics, University of Colorado (USA)}

%
%
%
%

\begin{abstract}
In preparation for the MMS mission we ask the question: how common are bipolar signatures linked to the presence of electron holes along separatrices emanating from reconnection regions? To answer this question, we conduct massively parallel simulations for realistic conditions and for the hydrogen mass ratio in boxes larger than considered in similar previous studies.
{ The magnetic field configuration includes both a field reversal and a out of plane guide field, as typical of many space situations. The guide field is varied in strength from low values (typical of the Earth magnetotail) to high values comparable to the in plane reconnecting field (as in the magnetopause). In all cases, along the separatrices a strong electron flow is observed, sufficient to lead to the onset of streaming instabilities and to form bipolar parallel electric field signatures. The presence of  bipolar structures at all guide fields allows the control of the MMS mission to consider the presence of bipolar signatures as a general flag of the presence of a nearby reconnection site both in the nightside and in the dayside of the magnetosphere.}

\end{abstract}

\begin{article}

\section{Introduction}
NASA's Magnetospheric Multi Scale Mission~\citep{mms1,mms2} will "use four identical spacecraft, variably spaced in Earth orbit, to make three-dimensional measurements of magnetospheric boundary regions and examine the process of magnetic reconnection" (quotation from the web site, mms.gsfc.nasa.gov). The authors are part of one of the theory teams that is entrusted with the task of exploring and quantify specifically the conditions to be expected for the mission.

One of the most important tasks is to predict, qualitatively and quantitatively,  what signatures are expected under realistic conditions for typical crossings of the satellites in regions of ongoing reconnection. We focus here on one specific diagnostics that is among those considered candidates to flag reconnection: the presence of bipolar signatures in the electric field. No single diagnostics is likely to be a viable flag of reconnection when used by itself, but in combination with others, the presence of bipolar electric structures can be a valuable signal to consider.

{ A key aspect of reconnection is the formation of strong electron flows. The electrons become decoupled from the ions and can reach speeds greatly exceeding the Alfv\'en speed, typical of the bulk plasma motion in the vicinity of a reconnection site. The relative drift of electrons and ions or between different electron populations can then largely exceed  the electron thermal speed and  streaming instabilities can develop and form bipolar structures~\citep{drake-science,goldman-holes}.

Electron holes are localised regions of reduced electron density that lead to a bipolar electrostatic field. Bipolar signatures in the parallel electric field  have indeed been observed in satellite crossings both in the magnetotail~\citep{cattell, pickett} and in the magnetopause~\citep{retino-holes}.


The analysis of the signatures observed by Cluster  \citep{cattell}  led to the conclusion that bipolar structures tend to be observed predominantly in the edge of the current sheets, along the separatrices (i.e. the field lines emerging from the center of a reconnection site).   While bipolar  signals the relative vicinity of reconnection regions, their presence is reported also at a distance of up to 50 ion inertial lengths along the separatrices~\citep{retino-holes}.

{ Simulations have also shown the presence of bipolar structures.  \citet{drake-science} considered an initial force free equilibrium that included an initially imposed relative drift between ions and electrons (of two times the electron thermal speed) and a guide field of approximately 5 times the reconnecting field. The formation of bipolar structures was observed in the vicinity of the reconnection region. Initial equilibria more closely resembling parameters observed in the magnetic tail, including smaller guide fields of 1/2 of the reconnecting field, have been reported  by 
\citep{cattell}. The presence of bipolar structures was observed along the separatrices leading to the reconnection region. However, at guide fields smaller than 1/2 of the reconnecting fields, the formation of bipolar structures was suppressed. Instead, the same authors observe that in the observations bipolar structures are present also in cases where the guide field is smaller.}

 \citet{cattell} suggest that the disagreement was due to limitations in the simulations that were conducted for  a reduced unphysical mass ratio ($m_i/m_e=100$), a very large unphysical  Alfv\'en speed ($c/v_A=20$) and a small box size ($25.6 \times 12.8 d_i$). The combined use of large electron scales and small boxes introduces strong boundary effects: the electron motion along the separatrices is blocked by the close proximity of the boundary that, when measured in the electron scales, is only $128 d_e$ away from the x-point.  \citet{cattell} proposed that future work revisit the issue reducing the proximity of the boundary. This can be obtained by restoring the physical mass ratio (and thereby making the simulation box larger when measured in electron scales) and by dicrectly increasing the box size. Both operations are rather costly, because more resolution is needed to resolve the smaller electron scale and because of the use of a larger simulation domain.}




We have recently developed a massively parallel implementation of the implicit particle in cell method, the code iPic3D \citep{ipic}.  iPic3D  showed excellent scaling on massively parallel computers (up to but not limited to 16,000 processors) and the ability to conduct simulations at the physical mass ratio $m_i/m_e=1836$ with  relatively large box sizes ($20 \times 10 d_i$), resolving the electron skin depth, $d_e=c/\omega_{pe}$ \citep{mmspop}. The simulations require a tremendous computational cost, made possible by a targeted allocation on NASA's supercomputers for the theory support of the MMS mission.

{
In the present paper, to address the fundamental need to allow the unimpeded electron motion along the separatrices for long distances, we extend our previous simulations reported in \citet{mmspop}
 by further doubling the box size ($40 \times 20 d_i$) in each direction while retaining the same accuracy. We  used also 4 times as many processors, for a total of 4096. With the combined use of physical electron scales and of larger simulation boxes, the electrons can now move freely for nearly $900 d_e$ before the boundary becomes an issue. }

The results of the study remove the previous inconsistency of simulations and observations. When the electron motion along the separatrices is unimpeded by the boundary conditions, strong bipolar signatures typical of electron holes are present at any guide field, with the same field intensities as observed in the satellite measurements.

This is a crucial result for the MMS mission: bipolar structures will not be limited to the cases when reconnection happens in presence of significant guide fields. Rather their presence is a universal signature of reconnection and is therefore a precious requisite for identifying a reconnection regions.


\section{Conditions  of the Guide Field Scaling Study}

We consider an initial Harris equilibrium,
with $ B_x=B_0 \tanh(y/L)$
and  $p=p_b+p_0 {\rm sech}^2(y/L)$,
where  $x$ is along the sheared component of the magnetic field (Earth-Sun direction in the Earth magnetosphere), $y$ in the direction of the gradients (north-south in the magnetosphere) and $z$ along the current (dawn-dusk in the magnetotail). Note the difference from the more common GSM coordinate system.

The chosen physical parameters for the present study are consistent with the observed properties of the geomagnetic tail:  electron thermal velocity $v_{the}/c=.045$,  ratio of electron and ion temperatures $T_i/T_e=5$,   current sheet thickness $L/d_i=0.5$ and  background pressure $p_b/p_0=.1$. The  mass ratio is physical, $m_i/m_e=1836$ (neglecting oxygen and other minority species). These choices correspond also to a realistic Alfv\'en speed $c/v_a = 300$~\citep{birn-book}.

All the simulations use a computational box with $L_x/d_i=40$ and $L_y/d_i=20$. Periodic boundary conditions are used in $x$ and Dirichelet boundary conditions ($\bfE_t=0$ and $\delta \bfB_n=0$) in $y$. The $z$ direction is assumed to be ignorable, but all fields and the velocities of the particles have also the $z$ component making the simulation 2D3V. During the evolution the electron and ion scales change dramatically and locally. We have recently studied the grid resolution required to resolve the local electron skin depth in the regions of interest where the bipolar structures develop \citep{mmspop}. The same identical  resolution in space and time is used here and the reader is referred to the previous study for a detailed discussion.

We have changed parametrically the guide field: $B_z/B_0=0,.05,.1,.25,.3,1$. Reconnection is initiated by a  localized perturbation \citep{mmspop} that allows the system to evolve naturally without further boundary driving. The simulation is continued while reconnection evolves, and it is stopped before the process starts to be affected by the boundary conditions and saturates.

\begin{figure}[ht]
\includegraphics[width=\columnwidth,angle=0]{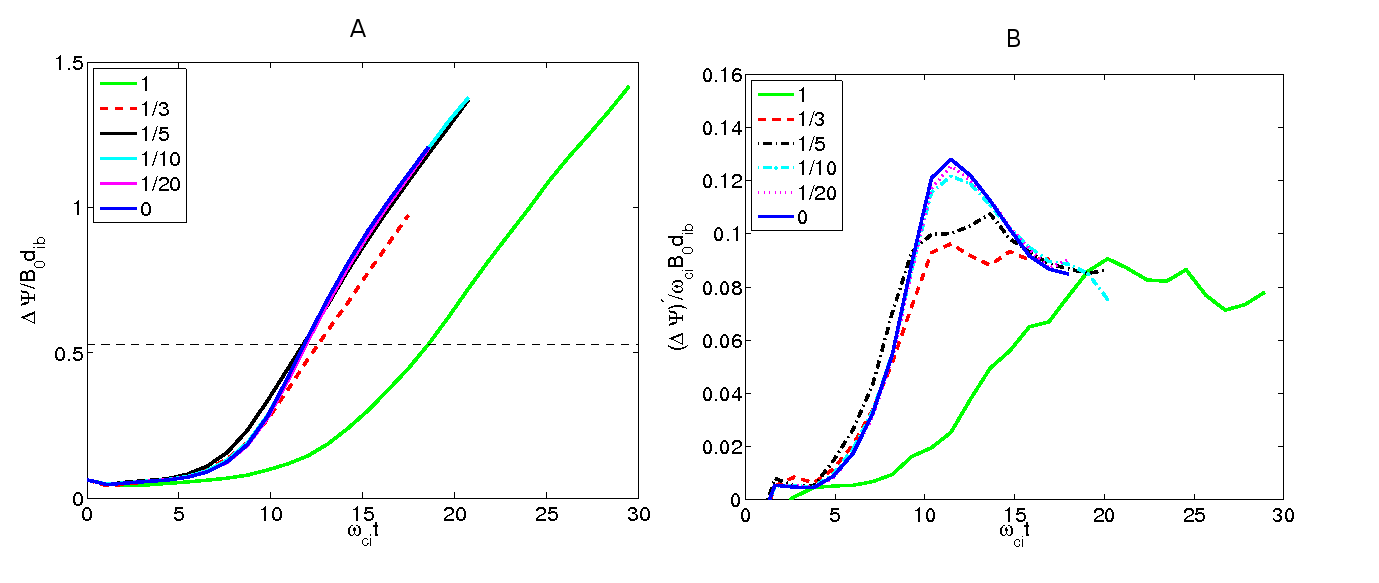}
\caption{{History of the reconnected flux (normalised using the in plane lobe magnetic field $B_0$ and density $n_b$, left) and reconnection rate (right) for  runs  at  different guide fields indicated in the legends. The horizontal dashed line reports the level of reconnected flux used in the analysis below.}} \label{recon}
\end{figure}

The resulting history of the reconnected flux and of the reconnection rate are reported in Fig.~\ref{recon}.
As previous studies have indicated~\citep{ricci-guide}, the reconnection rate remains strong for all guide fields, with a progressive decrease of the reconnection rate with increasing guide fields that becomes significant when the plasma $\beta$ in the guide field drops below unity. 
~\citep{rogers-waves,rogers-signatures}.

\begin{figure*}[ht]
\includegraphics[width=\textwidth,angle=0]{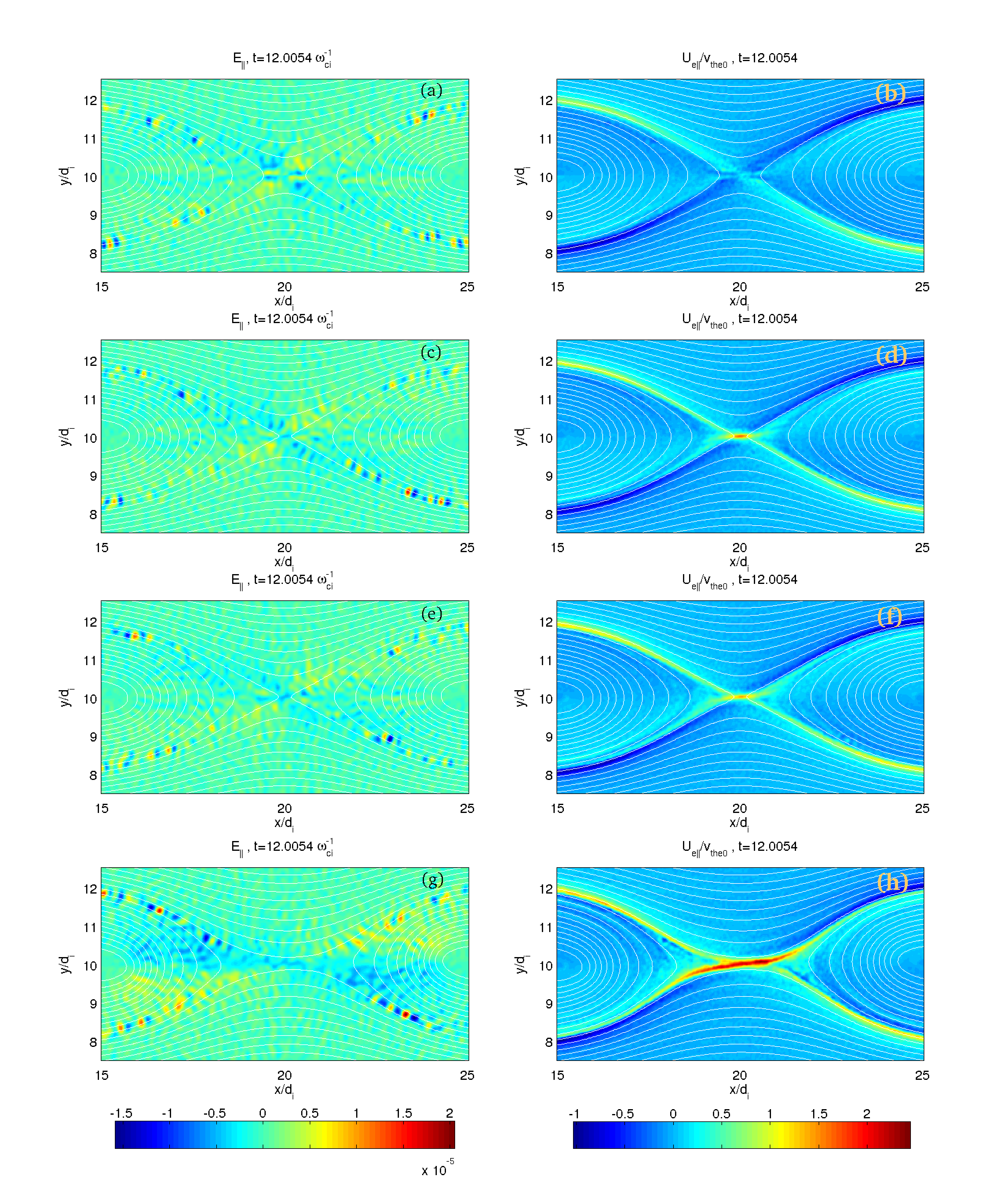}
\caption{Left: Parallel electric field, normalized as $eE_{||}/(m_ic\omega_{pi})$. Right: Parallel electron velocity normalized as $U_{e||}/v_{the}(x,y)$ (left). From top to bottom, guide fields: $B_z/B_0=0,.05,.1,.25$. Superimposed in all plots (in white): intersections of the magnetic field surfaces (contours of $A_z$). The time for each plot corresponds to approximately the same level of reconnected flux shown in Fig.~\ref{recon}: the actual time chosen is that of the closest graphics dump from the run. Here the closest graphics dump is the same for all runs. }.  \label{scaling-low}
\end{figure*}

\begin{figure*}[ht]
\includegraphics[width=\textwidth,angle=0]{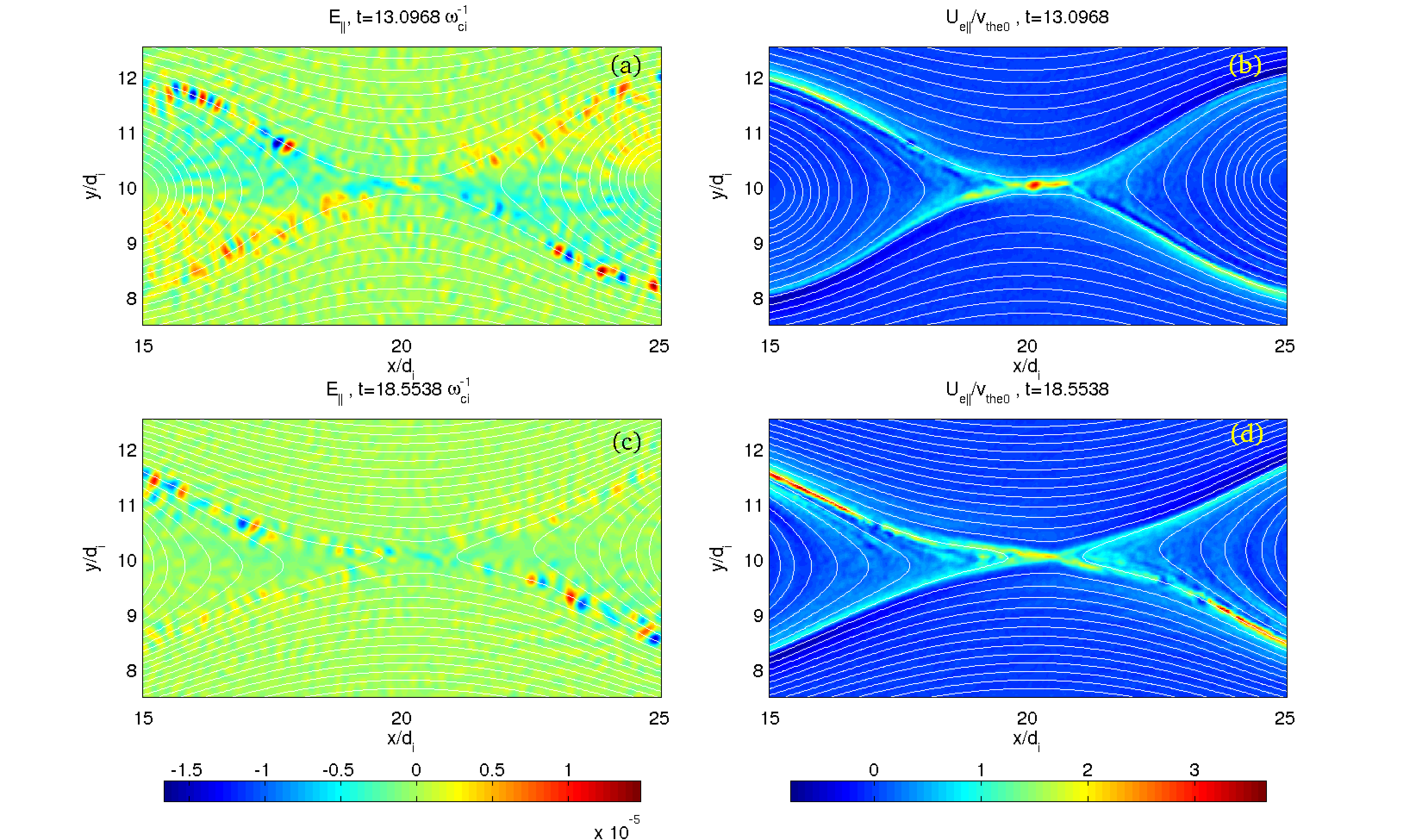}
\caption{Left: parallel electric field, normalized as $eE_{||}/(m_ic\omega_{pi})$. Right: Parallel electron velocity normalized as  $U_{e||}/v_{the}(x,y)$ (left). From top to bottom, guide fields: $B_z/B_0=1/3,1$. See Fig.~\ref{scaling-low} for details.} \label{scaling-high}
\end{figure*}

  Figure \ref{scaling-low} and  \ref{scaling-high} summarize the results of our investigation. The parallel electric field ($E_{||}={\bf E}\cdot {\bf B}/B$) is shown on the left and the corresponding parallel electron speed ($U_{e||}={\bf U}_e\cdot {\bf B}/B$) is shown on the right for each different guide field. Given the different rate of reconnection  for the different runs, we choose for each run a time corresponding  to the same reconnected flux (indicated in the figure by the horizontal line). {The choice of the time during the reconnection process is not critical. As reported in \citet{mmspop}, the electron flux and the bipolar signature are present with similar properties in all phases, at the peak of the reconnection rate (the time chosen in the present study) on later when it reaches a slightly lower value.} {Two movies of the typical evolution of one of the runs (with $B_z/B_0=.2$) is reported in the supporting material online, for the electron velocity and electric field parallel to the local magnetic field. }

{Two regions of strong electron flow can be identified: a) the electron diffusion region located right at the x-point with the electron jets ejected from it; b) the incoming separatrices where the electrons are sucked into the diffusion region.

The diffusion region and the outgoing jets are more prominent at low guide fields and are characterised by hot electrons with local thermal speeds comparable with the flow speed. Under these conditions streaming instabilities cannot easily develop while other weaker instabilities might be present~\citep{goldman-holes}. However, at the x-point and its immediate vicinity, the parallel electron flow is predominantly in the out of plane direction and its stability cannot be studied by the present 2D simulations. }

The present study, instead, focuses on the separatrices where the
electron flow has a large in-plane component and is the origin of the Hall current connected to the separation of scales in magnetic reconnection between electron and ions. This topic has been intensely investigated in the past and is reviewed in \citet[chapter 3]{birn-book}. As the guide field is increased, the pattern of field-aligned electron flow transitions from  antisymmetric (horizontally and vertically) to symmetric~\citep{rogers-signatures,swisdak-component}. At low guide fields, $U_{e||}$ is antisymmetric. Recalling that the upper and lower magnetic fields have opposite sign, this leads to a flow that is always moving along the separatrices inward towards the reconnection region. At higher guide fields, this pattern changes and the flow remains inward-directed only along the upper-left and lower-right separatrices (for the present choice of the sigh of the guide field). Instead, along the other two, it reverses and becomes outward directed. Additionally,  the inward-directed flow becomes significantly stronger than the outward flow along the other two separatrices.  { This is the reason why smaller boxes can pick up the formation of bipolar signatures only for strong guide fields: there the flow is stronger and the boundary conditions affect the electron physics less.}

In all cases, however, the flow speeds exceed the electron thermal speed and the strength increases as the guide field is increased. Note the different color scale in Fig. \ref{scaling-low} and  \ref{scaling-high}. The flow speed is so considerably larger in the highest guide field cases as to advise the use of a different scale for a better representation.

Under these conditions, streaming instabilities  lead to the formation of bipolar structures associated to regions of depleted electron density, referred to in the literature as electron holes~\citep{drake-science,goldman-holes}. The left column of  Fig. \ref{scaling-low} and  \ref{scaling-high} reports the observed bipolar structures as a function of the initial guide field. The electron flow along the separatrices is stronger and leads to the formation of bipolar parallel electric field signature.
In all cases, bipolar structures are present, and their strength is not affected significantly by the presence of guide fields. However, the predominant strength of the inward flow along the top-left and bottom-right separatrices tends to increase the prevalence of bipolar signatures along those flow channels without suppressing them altogether along the other two separatrices.

\section{Discussion}
The results above relate directly to the expected signatures to be observed by the MMS mission. The primary novelty resides on the choice of the physical parameters used: i) a large computational box, ii) the physical mass ratio for hydrogen and iii) a realistic value of the ratio of the  Alfv\'en speed with the speed of light. These three conditions  constitute a first. Previous studies were limited by computational resources to small mass ratios or to unphysically hot electrons or to small box sizes. Instead, the results presented here remove all these limitations thanks to the use of the massively parallel code iPic3D and thanks to a large computing time allocation on Pleiades devoted to the MMS mission.

The conclusion of the present study can be summarized in two main points.

First, we observe bipolar structures always, for all guide fields, including for zero guide field.  This is a key prediction of relevance to the MMS mission: bipolar signatures associated with electron holes are expected to be a common feature of reconnection separatrices, regardless of the intensity of the guide magnetic field, for component or antiparallel reconnection.
This extends previous findings that were able to find bipolar structures with guide fields from approximately 5 times~\citep{drake-science} to 1/2 times~\citep{cattell} of the reconnecting fields.


Second, the present study does quantitative rather than qualitative prediction thanks to the physical mass ratio and the realistic choice of other parameters. { The strength and size of the bipolar structures were studied in detail in \citep{mmspop} for the case of $B_z/B_0=0.5$ and it is found here to be insensitive to the guide field value for the range considered.}. The predicted values of the electron speeds and bipolar signature strengths are of primary importance. These can be directly compared with  satellite observations. { The bipolar signatures observed have a physical extension of approximately 16$\lambda_{De,loc}$ (where the Debye length is measured with the local density and temperature of the regions where the bipolar signatures are found) and are of the order  $eE_{||}/(m_ic\omega_{pi})=1.5\cdot 10^{-5}$. \citep{mmspop} These values are in good agreement with the observations reporting electric fields of the order of 10-30mV/m~\citep{cattell}, corresponding to $eE_{||}/(m_ic\omega_{pi})=(1-3)\cdot 10^{-5}$.}

The present study has spanned guide fields typical of both the magnetotail and the magnetopause but the latter presents also distinct asymmetries in the magnetic field and density between the solar and Earth side. This effect has been neglected here and future work will need to consider the effect in preparation of the first phase of the MMS mission devoted to the magnetopause. { Another important limitation is that the present study is 2D and flow instabilities can only develop along the separatrices. To study the close proximity of the x-point, where the flow is in the out of plane direction, full 3D runs are the next step.} This limitation is dictated by the need to use the correct physical mass ratio, a realistic Alfv\'en speed, sufficient accuracy and a large simulation box. These choices have been so far incompatible with a full 3D simulation. Yet the present runs used only 4096 cores of the 91,136 available on Pleiades. Newer supercomputers have even more. Our future work will focus on  full 3D study.

\begin{acknowledgments}
The present work is supported by NASA MMS Grant NNX08AO84G. Additional support for the KULeuven team is provided by the
Onderzoekfonds KU Leuven (Research Fund KU Leuven) and by
the European Commission's Seventh Framework Programme
(FP7/2007-2013) under the grant agreement no. 218816
(SOTERIA project, www.soteria-space.eu) and no. 263340 (SWIFF project, www.swiff.eu). The simulations were conducted on the resources of the NASA Advanced Supercomputing Division (NAS) and of the Vlaams
Supercomputer Centrum (VSC) at the Katholieke Universiteit
Leuven.
\end{acknowledgments}

\providecommand{\noopsort}[1]{}\providecommand{\singleletter}[1]{#1}%

\section*{Figure Captions}

Figure 1: History of the reconnected flux (normalised using the in plane lobe magnetic field $B_0$ and density $n_b$, left) and reconnection rate (right) for  runs  at  different guide fields indicated in the legends. The horizontal dashed line reports the level of reconnected flux used in the analysis below.

Figure 2: Left: Parallel electric field, normalized as $eE_{||}/(m_ic\omega_{pi})$. Right: Parallel electron velocity normalized as $U_{e||}/v_{the}(x,y)$ (left). From top to bottom, guide fields: $B_z/B_0=0,.05,.1,.25$. Superimposed in all plots (in white): intersections of the magnetic field surfaces (contours of $A_z$). The time for each plot corresponds to approximately the same level of reconnected flux shown in Fig.~\ref{recon}: the actual time chosen is that of the closest graphics dump from the run. Here the closest graphics dump is the same for all runs.

Figure 3: Left: parallel electric field, normalized as $eE_{||}/(m_ic\omega_{pi})$. Right: Parallel electron velocity normalized as  $U_{e||}/v_{the}(x,y)$ (left). From top to bottom, guide fields: $B_z/B_0=1/3,1$. See Fig.~\ref{scaling-low} for details.

\end{article}
\end{document}